\documentstyle[epsfig]{aipproc}

\begin{document}

\title{The Current Performance of the Third Interplanetary Network}
\author{K. Hurley$^1$, T.Cline$^2$, I. Mitrofanov$^3$, E. Mazets, S. Golenetskii$^4$, F. Frontera$^{5,6}$, E. Montanari, C.
Guidorzi$^6$, and M. Feroci$^7$ }
\address{$^1$UC Berkeley Space Sciences Laboratory, Berkeley, CA 94720-7450\\
$^2$NASA GSFC, Code 661, Greenbelt, MD 20771\\
$^3$IKI, 117810, Profsouznaya 84/32. GSP-7, Moscow, Russia\\
$^4$Ioffe Physical-Technical Institute, St. Petersburg, 194021, Russia\\
$^5$Istituto Tecnologie e Studio Radiazioni Extraterrestri, CNR, Via Gobetti 101, 40129, Bologna, Italy\\
$^6$Universita di Ferrara, via Paradiso, 12, 44100 Ferrara, Italy\\
$^7$Istituto di Astrofisica Spaziale, C.N.R., via Fosso del Cavaliere, Rome I-00133, Italy}

\maketitle

\begin{abstract}

The 3rd Interplanetary Network (IPN) has been operating since April 2001 with two distant spacecraft,
Ulysses and Mars Odyssey, and numerous near-Earth spacecraft, such as BeppoSAX, Wind, and HETE-II.
Mars Odyssey is presently in orbit about Mars, and the network has detected approximately 30 cosmic, SGR,
and solar bursts.  We discuss the results obtained to date and use them to predict the future performance of
the network.

\end{abstract}

\section{Introduction}

The 3rd IPN began with the launch of Ulysses in November 1990.  Ulysses is in a heliocentric orbit roughly perpendicular to the ecliptic, with perihelion  of about 1.5 AU and an aphelion of about 5 AU.  Until 1992, the network had Ulysses and Pioneer Venus Orbiter (PVO) as its distant points, and utilized many near-Earth spacecraft such as the Compton Gamma-Ray Observatory (CGRO) as its third point.  PVO entered the atmosphere of Venus in 1992, and it was to be replaced in the IPN by NASA's Mars Observer, which was lost during insertion into Martian orbit.  Finally, in 1999, the network was completed by the X- and Gamma-Ray Spectrometer experiment aboard the Near Earth Asteroid Rendezvous (NEAR) spacecraft.  In this configuration, the IPN operated quite successfully
until NEAR landed on the asteroid Eros in February 2001.  Figure 1 summarizes some of the results.
The Mars Odyssey mission, launched in April of that year, contains two experiments which
have been modified for gamma-ray burst detection, the Gamma-Ray Spectrometer (GRS) and the High Energy Neutron Detector (HEND).  As the GRS
will not be turned on permanently until the spacecraft completes its aerobraking maneuver, we will discuss the results obtained with HEND.

\begin{figure}
\psfig{file=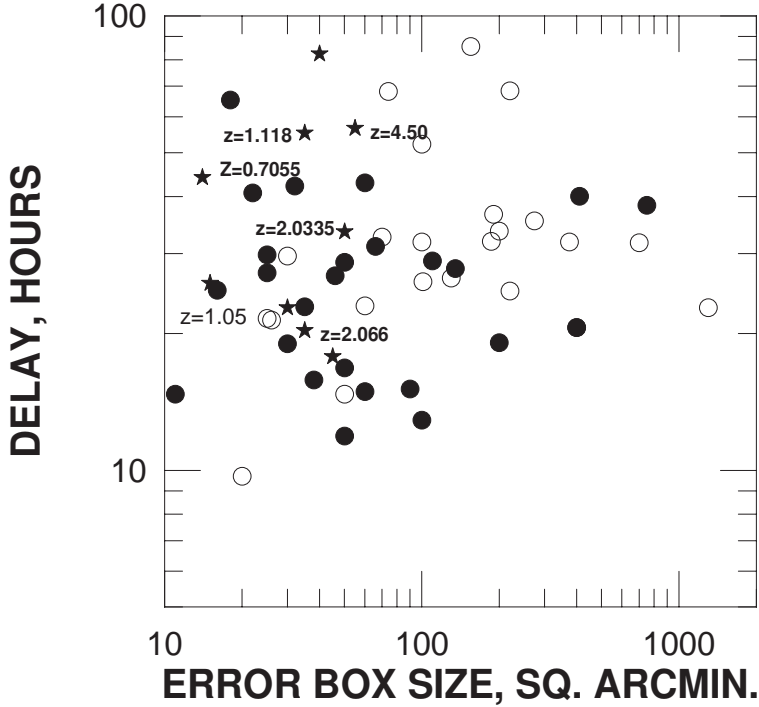,width=10cm}
\caption{Fifty-seven GRBs detected by the IPN with NEAR.  Each burst is characterized by the size of the
error box and the delay to obtain it.  The stars represent GRBs for which counterparts were identified; six of them
have had their redshifts measured.  The filled circles represent bursts which were followed up in the radio and/
or optical range, but for which no counterparts were identified.  The hollow circles represent those events
which were not followed up.  In slightly over a year of operation, the database on GRB counterparts
increased by 50\% due to the IPN results alone.}
\label{fig1}
\end{figure}

\section{Some IPN results to date}

HEND has an effective area of about 40 cm$^2$ of CsI and operates in the energy range above 40 keV.  It transmits data
continuously with a time resolution of 0.25 s, so that "triggering" is done with ground software.  In addition, whenever a burst
is recorded by Konus, the BeppoSAX GRBM, Ulysses, or HETE-II, HEND data are extracted for the appropriate crossing
window and transmitted to Goddard and Berkeley for analysis.  HEND was been on for a large fraction of the cruise phase, and has detected many cosmic, solar, and soft gamma repeater (SGR) events, some of which are listed in table 1. 
\begin{table}[!t]
\begin{tabular}{cccccccc} 
\hline
Date & Seconds & Type & HEND & Ulysses & HETE &  Konus & SAX \\
\hline

010508	&47828	&Cosmic	&Yes	& Yes	& No	& Yes	& No \\
010517	&85894	&Cosmic	&Yes	& Yes	& No	& Yes	& Yes \\
010523	&17059	&Cosmic	&Yes	& Yes	& No	& Yes	& No \\
010605	&17145	&Solar	           & Yes	 & No	& No	& Yes	& No \\
010607	&53723	&Cosmic	& Yes	& Yes	& No	& Yes	& No \\
010625	&51802	&SGR1900+14	& Yes	& Yes	& No	& Yes	& No \\
010627	&04172	&Cosmic	& Yes	& Yes	& No	& No	& No \\
010628A	&03914	&Cosmic	& Yes	& Yes	& Yes	& Yes 	& No \\
010628B	&68418	&Cosmic	& Yes	& Yes	& No	& Yes	& No \\
010701A	&02789	&Cosmic	& Yes	& Yes	& No	& Yes	& No \\
010701B	&07102	&Cosmic	& Yes	& Yes	& No	& Yes	& No \\
010702	&12848	&SGR1900+14	& Yes	& Yes	& Yes	& Yes	& Yes \\
010703	&73842	&Cosmic	& Yes	& Yes	& No	& Yes	& Yes \\
010706	&29689	&Cosmic	& Yes	& Yes	& No	& Yes	& No \\
010710	&84642	&Cosmic	& Yes	& Yes	& No	& Yes	& Yes \\
010723	&63602	&Cosmic	& Yes	& Yes	& No	& Yes	& No \\
010725	&61288	&Cosmic	& Yes	& Yes	& No	& Yes	& No \\
010726	&05392	&Cosmic	& Yes	& Yes	& No	& Yes	& No \\
010804	&72805	&Cosmic	& Yes	& Yes	& No	& Yes	& Yes \\
010805	&54213	&Solar 	& Yes	& Yes	& No	& Yes	& No \\
010807	&59183	&Solar	& Yes	& Yes	& No	& Yes	& No \\
010821A	&48423	&Cosmic	& Yes	& Yes	& No	& Yes	& No \\
010821B	&78900	&Solar 	& Yes	& Yes	& No	& Yes	& No \\
010827	&23676	&Solar	& Yes	& Yes	& No	& Yes	& No \\
010830	&74159	&Solar	& Yes	& Yes	& No	& Yes	& No \\
010831A	&38298	&Solar	& Yes	& Yes	& No	& No	& No \\
010831B	&81626	&Solar	& Yes	& Yes	& No	& Yes	& No \\
010913	&71191	&Cosmic	& Yes	& Yes	& No	& Yes	& No \\
010921	&18552	&Cosmic	& Yes	& Yes	& Yes	& Yes	& Yes \\
\hline
\end{tabular}
\caption{A partial list of HEND solar, cosmic, and SGR bursts.}
\label{tab:b}
\end{table}

This table is neither exhaustive, since we have not been able to analyze all the data yet, nor representative, since there were numerous solar flare particle events which raised the backgrounds and reduced the sensitivities of some IPN experiments. (Solar activity will be decreasing in the years to come.)  Thus in about 4.5 months, HEND detected 21 confirmed cosmic or SGR events.  HEND is now on almost continuously, but in an eccentric orbit which leads to a variable background rate.
The orbit is gradually being circularized by aerobraking. In figure 1 we show an example of a HEND burst.  Although the GRS has been on only sporadically during cruise, it too has detected several bursts.

\begin{figure}
\psfig{file=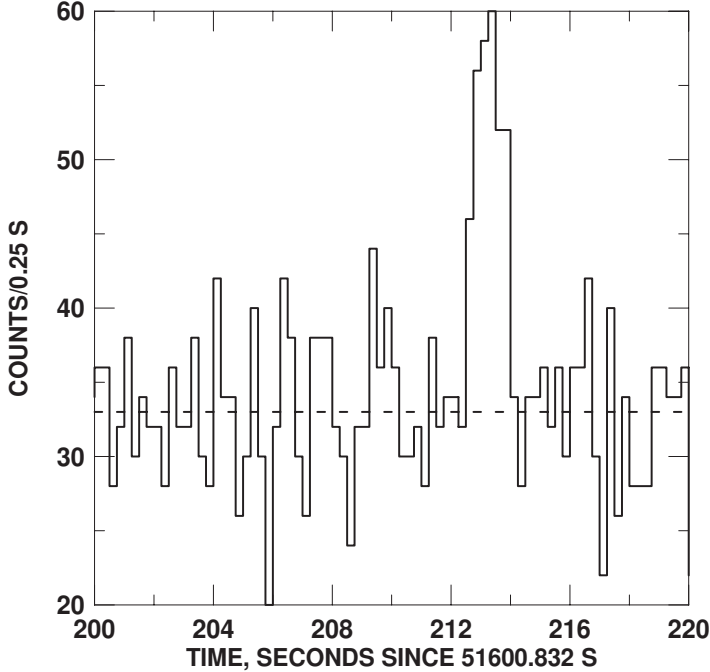,width=10cm}
\caption{The time history of GRB010625 as observed by HEND in the 50-3000 keV energy range.  The dashed line shows the background.
This event was from SGR1900+14 and had a fluence of approximately $\rm 3 \times 10^{-6} erg cm^{-2}$}
\label{fig2}
\end{figure}

\section{The future of the IPN}

The Mars Odyssey mission will collect data for at least one Martian year.  The nominal end of mission will be around
early 2004.  Ulysses is currently funded through at least the start of 2004, with a slight extension possible.  However,
the decay of the radioactive power system probably precludes operation past the end of 2004.  The Wind mission is funded through
2002 at least.  The BeppoSAX is presently funded until April 2002, but it may be extended until it re-enters.  HETE-II is funded through 2002 at least, and
possibly for two years beyond.

The INTEGRAL mission is due to be launched into a highly elliptical orbit in October 2002.  It will therefore either replace or complement the
near-Earth spacecraft, depending on whether they remain operational or have been terminated by that time.  Finally, Swift should
be launched in late 2003.  Thus, we expect to maintain a viable three-spacecraft IPN for a minimum of about 2.5 years. 

\section{Expected results}

This new IPN is similar in many respects to one which was in operation when NEAR was in the network.  Between December 1999 and January 2001, that IPN localized 57 GRBs and circulated their positions to the wide astronomical community via the GRB Coordinates Network (GCN).  Of the 25 bursts which were followed up with long-wavelength observations, counterparts were found for 9.  The success rate, 9/25 or 36\%, is consistent with the overall success rate for bursts, which is about 40\%.  

There are factors which will both increase and decrease the performance of the new IPN compared to the old one.  We first list the "increase" factors.

1. In the old network, the Earth-Ulysses distance varied between 2.3 and 4.4 AU, while the Earth-NEAR distance varied between 0.8 and 2.1.  In the new network, the Earth-Ulysses distance will vary between 2 and 6 AU, while the Earth-Mars distance will vary between 0.45 and 1.8 AU.  Thus the average baselines, and hence the annulus widths, will be slightly better.

2. In the old network, the NEAR XGRS experiment had a time resolution of 1 second.  In the new network, the time resolution will be between 0.031 s (for
GRS) and 0.250 s (for HEND), resulting in a gain in accuracy of the cross-correlations and thus narrower annuli.

3. In the old network, the XGRS detector had a lower energy threshold of ~150 keV.  The other experiments in the network have energy ranges of ~25-150 keV.  Since GRB time histories are energy-dependent, this led to increased uncertainty in the cross-correlations and thus wider annuli.  HEND operates in an energy range which is better matched to those of the other IPN instruments.

4. Based on the number of bursts observed with HEND during the cruise phase (21 in 4.5 mo), and taking into account the facts that HEND was not on continuously, and that all the data have not been examined, the burst detection rate, ~4.7/month, was greater than that of the old network.  

The following are "decrease" effects:

1. Once MO is in a circular orbit, HEND will have an unobstructed field of view of about 2 pi sr, or a factor of about 2 less than in
the cruise phase.  This decrease should be offset by two other factors, however.  First, bursts arriving from the anti-planet hemisphere should still be detectable in some cases, since they will either penetrate the material behind the detectors, and/or backscatter off the Martian atmosphere.  Second, GRS will be on continuously, and due to its different detection efficiency, energy range, and field of view, it will detect some bursts that HEND does not.

2. The future tracking efficiencies (that is, the length of time between telemetry passes) of some spacecraft in the network are unknown.  
This applies particularly to Konus-Wind and to Mars Odyssey.  Longer periods between downlinks translate to later error boxes, which
must be searched more deeply for counterparts.

All these effects are difficult to quantify, but our expectation is nevertheless that the performance of this network will closely resemble
that of its predecessor.

\section{Conclusions}

In the 2.5 year MO mission, we expect to detect and localize over 100 GRBs and circulate their positions rapidly to astronomers around the world for multiwavelength counterpart studies.  This should lead to the observation of numerous afterglows and the determination of many GRB redshifts.

KH is grateful for IPN support under JPL Contract 958056.

\end{document}